%% file: main.tex
\documentclass[9pt]{extarticle}
\usepackage[letterpaper,margin=2.2cm]{geometry}
\usepackage{graphicx}
\graphicspath{{figures/}}
\usepackage[utf8]{inputenc}
\usepackage{booktabs}
\usepackage{array}
\usepackage{amsmath}
\usepackage{microtype}
\usepackage{titlesec}
\usepackage{caption}
\usepackage{enumitem}
\usepackage{placeins}
\usepackage[hidelinks]{hyperref}
\usepackage[htt]{xurl}
\titlespacing*{\section}{0pt}{2.1ex plus .3ex minus .2ex}{1.0ex plus .1ex}
\titlespacing*{\subsection}{0pt}{1.6ex plus .25ex minus .1ex}{0.7ex plus .1ex}
\titlespacing*{\subsubsection}{0pt}{1.4ex plus .2ex}{0.6ex plus .1ex}
\titlespacing*{\paragraph}{0pt}{1.1ex plus .2ex}{0.5em}
\setlist{topsep=0.25em,itemsep=0.12em,parsep=0pt,partopsep=0pt}
\title{SkillSeam: Six Principles\\for Auditing Agent Skill Collections}
\author{Kang Ruiyuan \\ \small X32 Studio, Abu Dhabi, UAE \\ \small \texttt{kang@x32studio.com}}
\date{}

\begin{document}
\maketitle

\input{figures/captions}% one-sentence figure captions (takeaways)
\input{sections/abstract}
\input{sections/intro}
\input{sections/method}

\input{sections/results}
\input{sections/principles}
\input{sections/discussion}
\input{sections/related}
\input{sections/limits}
\input{sections/conclusion}
\appendix
\input{sections/appendix-repro}

\bibliographystyle{unsrt}
{\small
\bibliography{references}
}

\end{document}

%% file: figures/captions.tex
% Centralized figure captions for the SkillSeam paper.

\newcommand{\figcaphero}{The organizational metaphor behind SkillSeam. The same capabilities appear on both sides; only their relationships change. Six principles turn conflicting roles and tangled information paths into layered knowledge, explicit gates, unique ownership, observable handoffs, and right-sized responsibilities.}

\newcommand{\figcapprincipleevidence}{SkillSeam separates context burden (P1--P2), routing and ownership failures (P3--P5), and task correctness (P6). P3 holds skill count and character budget fixed; P4 is a candidate-ownership diagnostic, not a natural-task failure rate.}

\newcommand{\figcapconceptprinciples}{Where SkillSeam acts in a skill system. P1--P2 govern persistence and traceability across layers; P3--P5 govern partitioning and routing among skills; P6 governs the grain of each skill.}

\newcommand{\figcapcontrolledprobes}{Mechanism-isolating P3/P4 probes. P3 replaces an equal-length unrelated control with a same-regime alias while holding the canonical skill fixed. P4 holds skill and section counts fixed while changing only the ownership matrix.}

%% file: sections/abstract.tex
\begin{abstract}
A folder of competent skills is not yet a reliable system. Skills
rarely fail alone; they fail at the seams of a collection. As an agent's
skill library grows, procedures compete for attention, aliases double-load,
boundaries blur, and poorly sized skills turn routing errors into task
failures. We introduce \textbf{SkillSeam}, a method that audits the
relationships through which individual skill files become a system.
It maps each collection-level principle to a failure mechanism, its
strongest observable, and a controlled perturbation test. From one
sealed skill system, SkillSeam
perturbs six design principles: persistence gradient, system
coherence, regime gating, orthogonal coverage, flow, and granularity
discipline. Crucially, each principle is evaluated through the channel
its failure mechanism predicts rather than through accuracy alone.
Flattening the persistence hierarchy increases loaded-skill tokens by
60\%; a dangling anchor raises total tokens by 64\% and shifts accuracy
by $-3.1$pp; with skill count and context size held fixed, replacing an
unrelated control with a synonymous alias raises noncanonical routes
from $0/32$ to $15/32$ and flips half of matched paraphrase pairs; in a
candidate-ownership audit, overlapping lanes raise reported ownership
conflicts from $0/16$ to $14/16$; bland triggers drive routing conflicts
from $3/32$ to $30/32$ and inflate loaded-skill tokens 3.7$\times$; and one
granularity mis-mix produces the largest accuracy drop, $-12.5$pp.
These outcomes turn six pieces of authoring advice into testable system
properties without treating every probe as confirmation. We release the byte-differenced variants, task
slices, rollups, and a one-screen design checklist so that other skill
systems can measure the same failure channels.
\end{abstract}

%% file: sections/intro.tex
\section{Introduction}
\label{sec:intro}

Giving an agent another skill resembles hiring another specialist: it
expands capability, but it does not create an organization. A roster
needs explicit roles, information layers, and handoffs; a skill
collection needs routing boundaries, ownership, traceability, and
testable contracts. Individual skill files are now a standard way to
give agents procedural knowledge~\cite{anthropic2025skills}. The harder
problem begins after the first few files: two skills claim the same
regime, a boundary is written where the router cannot observe it, a
reference becomes unreachable, or one skill absorbs several jobs. Each
skill may look competent alone while the collection fails at its seams.

\begin{figure*}[h]
\centering
\includegraphics[width=1.0\linewidth]{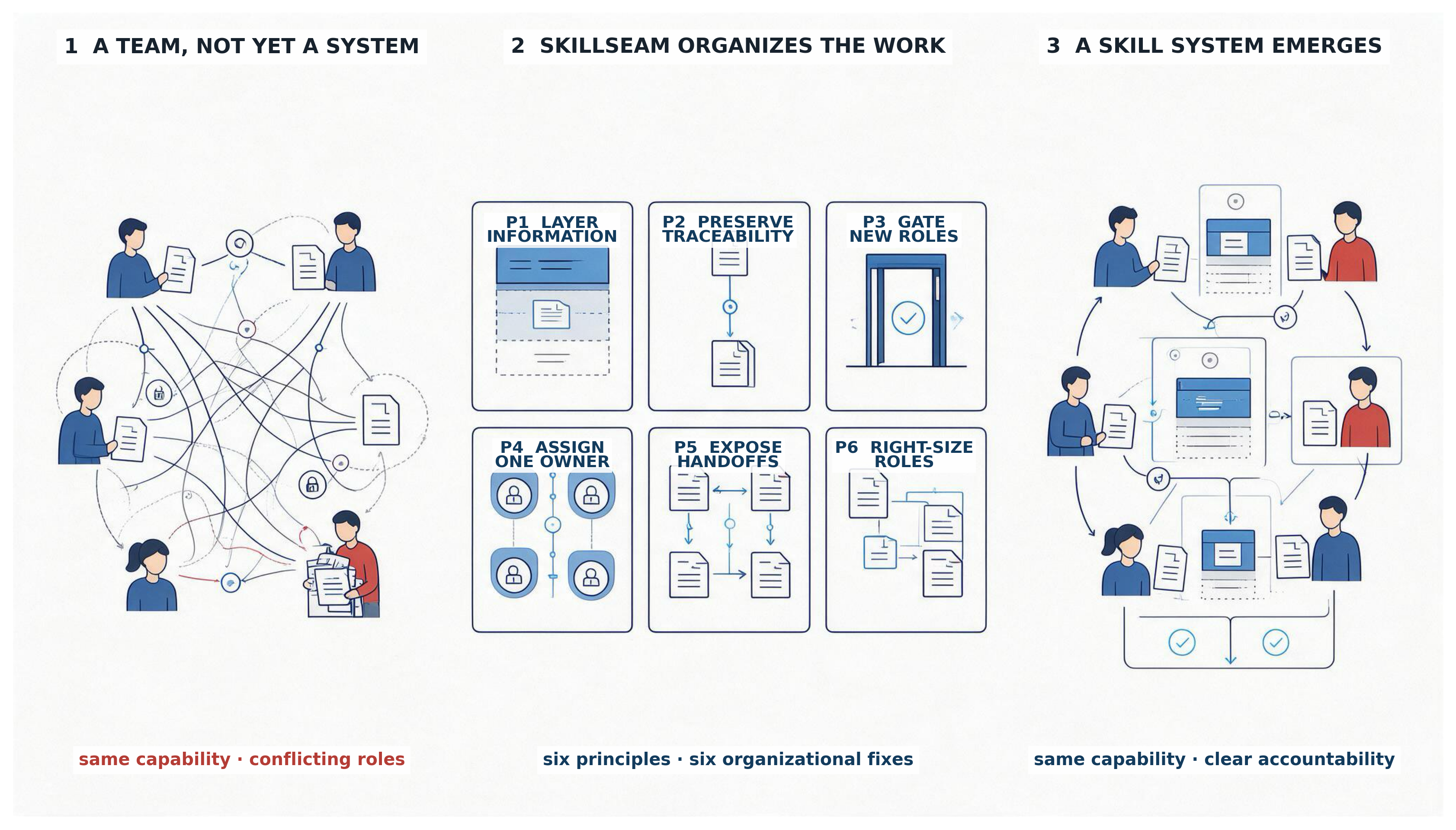}
\caption{\figcaphero}
\label{fig:hero}
\end{figure*}

Figure~\ref{fig:hero} is a controlled organizational thought experiment.
The people, skill cards, and underlying capability remain fixed; only
their relationships change. On the left, duplicate roles, broken
information paths, and implicit handoffs turn capable specialists into
a tangled organization. SkillSeam applies six interventions: layer
information, preserve traceability, gate new roles, assign one owner,
expose handoffs, and right-size responsibilities. The result is not a
smaller library. It is the same capability organized into a skill
system with clear accountability.

We introduce \textbf{SkillSeam} to ask a narrow question with broad
practical consequences:
\emph{what organizational relationships must remain decidable as a
roster of skills becomes a skill system?}
We study six collection-level principles---persistence gradient, system
coherence, regime gating, orthogonal coverage, flow, and granularity
discipline---derived from the System-of-Skills practitioner
guide~\cite{kr33skills}, which also ships the method as an installable,
agent-readable skill. Their common premise is that a skill collection is
an architecture: it has layers, interfaces, routing predicates, and
module boundaries. The principles should therefore be testable as
system properties, not defended as writing style.

Whether task-relevant procedures can help an agent is not our question;
our contribution begins after a team has committed to a skill layer.
We ask what breaks when that team organizes its skills badly, and we
measure each break in the channel where it should first appear.
Accuracy is appropriate for a mis-granulated capability; it is a weak
instrument for a routing collision. Loaded-skill tokens reveal
persistence and flow overhead; route flips reveal aliases; ownership
audits reveal overlaps; reachability and task success reveal coherence
and grain.

We make three contributions through SkillSeam:
\begin{enumerate}[leftmargin=1.6em]
  \item \textbf{A failure model for skill collections.} Each of six
  principles is paired with a concrete mechanism and an observable:
  context burden, unreachable traces, semantic multi-reads, overlap
  multi-reads, routing conflicts, or task failure.
  \item \textbf{A controlled perturbation method.} We generate a
  byte-differenced L0--L6 ladder from one sealed collection, changing one
  organizational property per rung, and add diagnostic slices where a
  failure cannot appear on positive tasks.
  \item \textbf{Measured mechanism-specific signatures.} The
  experiments expose effects ranging from 60\% more loaded-skill tokens
  to alias-driven route fragmentation in $15/32$ tasks, overlap
  conflicts in $14/16$ ownership audits, a $27/32$ increase in routing
  conflicts, and a $-12.5$pp accuracy drop.
  Table~\ref{tab:checklist} converts these outcomes into an executable
  audit.
\end{enumerate}

Figure~\ref{fig:hero} summarizes the argument: six principles protect
six different failure channels. Underneath them is self-contained
verification---a runnable test for every skill---whose specification
semantics we adopt from SkillSpec~\cite{zhang2026skillspec}. It is the
measurement substrate, not a seventh principle. The rest of the paper
defines the contrasts, reports the signatures, and turns them into a
design discipline that other agent systems can directly falsify.

%% file: sections/method.tex
\section{Measurement Design}
\label{sec:method}

SkillSeam is organized around one rule: measure each principle in
the channel where its predicted failure should become visible. A single
accuracy score cannot distinguish wasted context from double routing or
an unreachable dependency. We therefore predefine four outcome
families: task correctness, loaded-skill token burden, routing events
(conflict and multi-read), and execution effort (turns, tool calls, and
read calls).

\subsection{Harness, suite, and scoring}
A fixed Python/Node harness holds model configuration, system prompt,
temperature, injected seed, and task contract constant within every
matched comparison. The cumulative ladder uses temperature 0.7; the
targeted P3/P4 probes use GLM-5.3-Flash at temperature 0.2. Each session
produces a typed JSONL record containing task
correctness, selected and read skills, routing events, approximate token
counts, tool calls, and an invalidity reason. Accuracy is reported over
valid sessions; valid denominators and all timeout, refusal, and empty
responses remain in the released rollups.

The fixed 100-task suite spans communications, expenses, files, health,
and scheduling. Tasks are partitioned into positive, hard-negative,
ambiguous, out-of-domain, and paraphrase slices. Positive tasks test
whether the expected skill completes a verifiable workspace effect.
Hard negatives expose semantic aliases; ambiguous tasks expose overlap;
out-of-domain tasks test abstention. This partition matters because an
organizational defect is observable only on a task where its failure
condition can occur.

\subsection{Controlled principle perturbations}
The core experiment starts from one sealed, disciplined N16 collection
(L0). A generator creates six cumulative snapshots, each adding one
minimal organizational violation while preserving all untouched bytes.
The adjacent contrast $L_{k-1}\to L_k$ therefore estimates the change after
introducing the next violation in the accumulated system; adjacent
effects are not assumed independent or additive.

\begin{table}[h]
\centering\small
\caption{The controlled ladder. Each perturbation targets a distinct
failure mechanism and is read through its most sensitive observable.}
\label{tab:ladder-design}
\begin{tabular}{@{}clp{5.2cm}p{4.2cm}@{}}
\toprule
 & Principle & Perturbation & Primary observable \\
\midrule
P1 & Persistence gradient & Inline referenced detail into skill bodies & loaded-skill tokens \\
P2 & System coherence & Dangle one trace anchor & tokens, execution effort, accuracy \\
P3 & Regime gating & Replace an equal-length unrelated control with a same-regime alias & noncanonical route, paraphrase route flip \\
P4 & Orthogonal coverage & Make two equal-sized lanes claim the same operations & candidate ownership conflict \\
P5 & Flow & Replace yes/no triggers with bland labels & routing conflict, loaded-skill tokens \\
P6 & Granularity discipline & Merge one thin skill and split one coarse skill & task accuracy \\
\bottomrule
\end{tabular}
\end{table}

L0--L6 use 32 matched tasks per layer. P3 and P4 require separate
controlled diagnostics because their failures are structurally absent
from a positive-only slice. P3 uses 32 tasks arranged as 16
same-intent canonical/synonym pairs over four expense intents. Both
arms mount two skills and have the same character budget: the treatment
replaces an unrelated control with a synonym alias while leaving the
canonical skill byte-identical. P4 uses two skills with two sections
each in both arms. The baseline assigns two target operations to
disjoint owners; the treatment duplicates both claims. Sixteen balanced
candidate audits require an explicit
\texttt{OWNERSHIP\_CONFLICT: yes|no} decision. These probes are routing
diagnostics, not adjacent accuracy effects. We restrict every comparison
to sessions valid in both arms and retain a separate 32-task
direct-execution P4 slice to distinguish diagnostic sensitivity from
natural-task incidence.

\begin{figure}[h]
\centering
\includegraphics[width=0.98\textwidth]{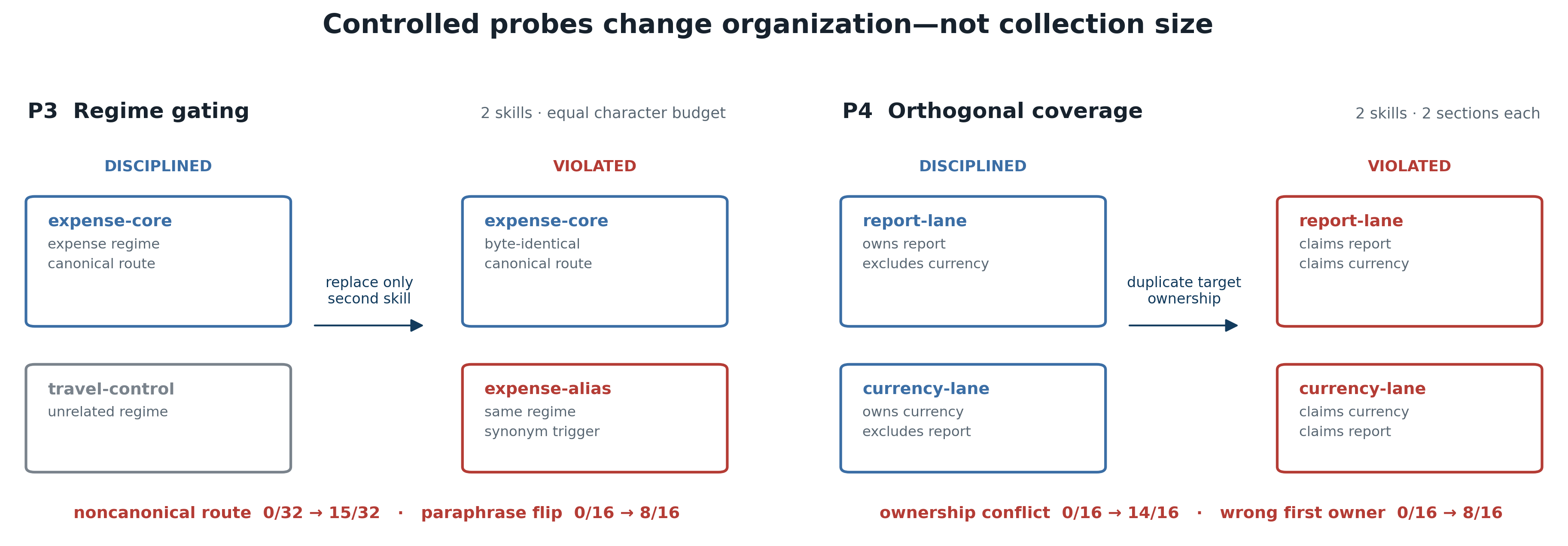}
\caption{\figcapcontrolledprobes}
\label{fig:controlled-probes}
\end{figure}

\subsection{Identification and alternative explanations}
Figure~\ref{fig:controlled-probes} makes the identifying contrasts
explicit. P3 could otherwise be dismissed as a collection-size or
prompt-length effect: adding any second file may create another read
candidate and consume more context. We therefore keep two files in both
arms, preserve \texttt{expense-core} byte-for-byte, and pad the
unrelated control to the exact character length of the alias. The 32
requests form 16 semantic pairs. Within each pair, arguments and desired
operation are fixed; only canonical versus synonymous wording changes.
A route flip is thus a change in which same-regime file becomes primary,
not a change in task semantics or available coverage.

P4 requires a different control. File count cannot isolate
orthogonality because the defect is a many-to-many ownership relation.
Both arms therefore contain the same two named lanes and two sections
per lane. The disciplined arm assigns report export and currency
conversion to separate owners, while the violated arm makes both files
claim both operations. Candidate audits expose whether the router can
name one owner after reading the available claims. We require a
machine-readable final decision,
\texttt{OWNERSHIP\_CONFLICT: yes|no}, and score only pairs with valid
decisions in both arms.

These probes intentionally optimize mechanism sensitivity. To prevent
that choice from being mistaken for an ecological rate, SkillSeam keeps the
earlier null target-pair probe and reports a separate direct-execution
P4 slice. The diagnostic answers whether overlap makes ownership
undecidable when inspected; the direct slice estimates how often an
ordinary task causes that defect to surface without an audit. Their
different magnitudes are part of the result, not interchangeable
replications.

\subsection{Supporting matrix}
A supporting mounting matrix contains 1,100 sessions across 11
conditions: managed and unmanaged layouts at N$\in\{4,16,64\}$, an N0
control, two N16 ablations, and a two-arm deepseek-v4-flash breadth
check. It serves three secondary purposes: verifying that the skill
layer is behaviorally consequential, checking whether skill count alone
explains performance, and showing why generic success tasks can conceal
P3/P4 routing failures.

\subsection{Analysis}
For paired binary outcomes we report adjacent percentage-point changes
and exact two-sided McNemar tests. The 32-task ladder is designed to
identify failure signatures, not to turn small accuracy differences
into significance claims; accordingly, the paper treats accuracy
deltas as effect estimates. The P5 routing contrast is the exception in
strength: 27 matched tasks switch from no conflict to conflict and none
switch in the opposite direction (exact $p=1.49\times10^{-8}$).
Continuous execution measures are summarized on matched valid tasks.
Token values are parser estimates, so we report relative token burden
rather than dollar cost or latency. The analysis script in
Appendix~\ref{sec:repro} recomputes every primary observable directly
from the JSONL rollups and probe reports. Exact paired tests give
$p=6.10\times10^{-5}$ for P3 noncanonical routing and
$p=1.22\times10^{-4}$ for P4 reported ownership conflict.

%% file: sections/results.tex
\section{Evidence of Organizational Failure}
\label{sec:results}

Table~\ref{tab:principle-evidence} is SkillSeam's central result. The six
perturbations do not produce six interchangeable accuracy deltas; they
produce six mechanism-specific outcomes. This is the point of the
measurement design: a test supports a principle only when breaking it
creates the predicted failure in the designated observable channel.

\begin{table}[h]
\centering\footnotesize
\setlength{\tabcolsep}{3pt}
\caption{Six principles, six failure mechanisms, and their strongest
measured outcomes. Token counts are approximate parser estimates;
P3/P4 use controlled routing diagnostics; other ladder comparisons are
adjacent.}
\label{tab:principle-evidence}
\begin{tabular}{@{}c>{\raggedright\arraybackslash}p{2.5cm}
>{\raggedright\arraybackslash}p{3.4cm}
>{\raggedright\arraybackslash}p{3.1cm}
>{\raggedright\arraybackslash}p{5.2cm}@{}}
\toprule
 & Principle & Failure mechanism & Primary channel & Observed signature \\
\midrule
P1 & Persistence gradient & Context carried at the wrong layer & loaded-skill tokens & $357\to570$ per matched valid session ($+59.5\%$); accuracy did not fall \\
P2 & System coherence & Broken trace creates search and recovery & total tokens / accuracy & $13{,}052\to21{,}400$ tokens ($+64.0\%$); $-3.1$pp accuracy estimate \\
P3 & Regime gating & Synonymous aliases split one semantic route & noncanonical route / paraphrase flip & routes $0/32\to15/32$ ($p=6.10{\times}10^{-5}$); flips $0/16\to8/16$ \\
P4 & Orthogonal coverage & Overlapping lanes create duplicate ownership & candidate ownership conflict & $0/16\to14/16$ ($p=1.22{\times}10^{-4}$); wrong first owner $0/16\to8/16$ \\
P5 & Flow & Non-observable triggers collapse routing & conflict / loaded-skill tokens & conflicts $3/32\to30/32$; tokens $552\to2{,}061$ ($3.7\times$) \\
P6 & Granularity discipline & Mis-sized modules fail task contracts & accuracy & $0.906\to0.781$ ($-12.5$pp), the largest adjacent accuracy change \\
\bottomrule
\end{tabular}
\end{table}

\begin{figure}[h]
\centering
\includegraphics[width=0.96\textwidth]{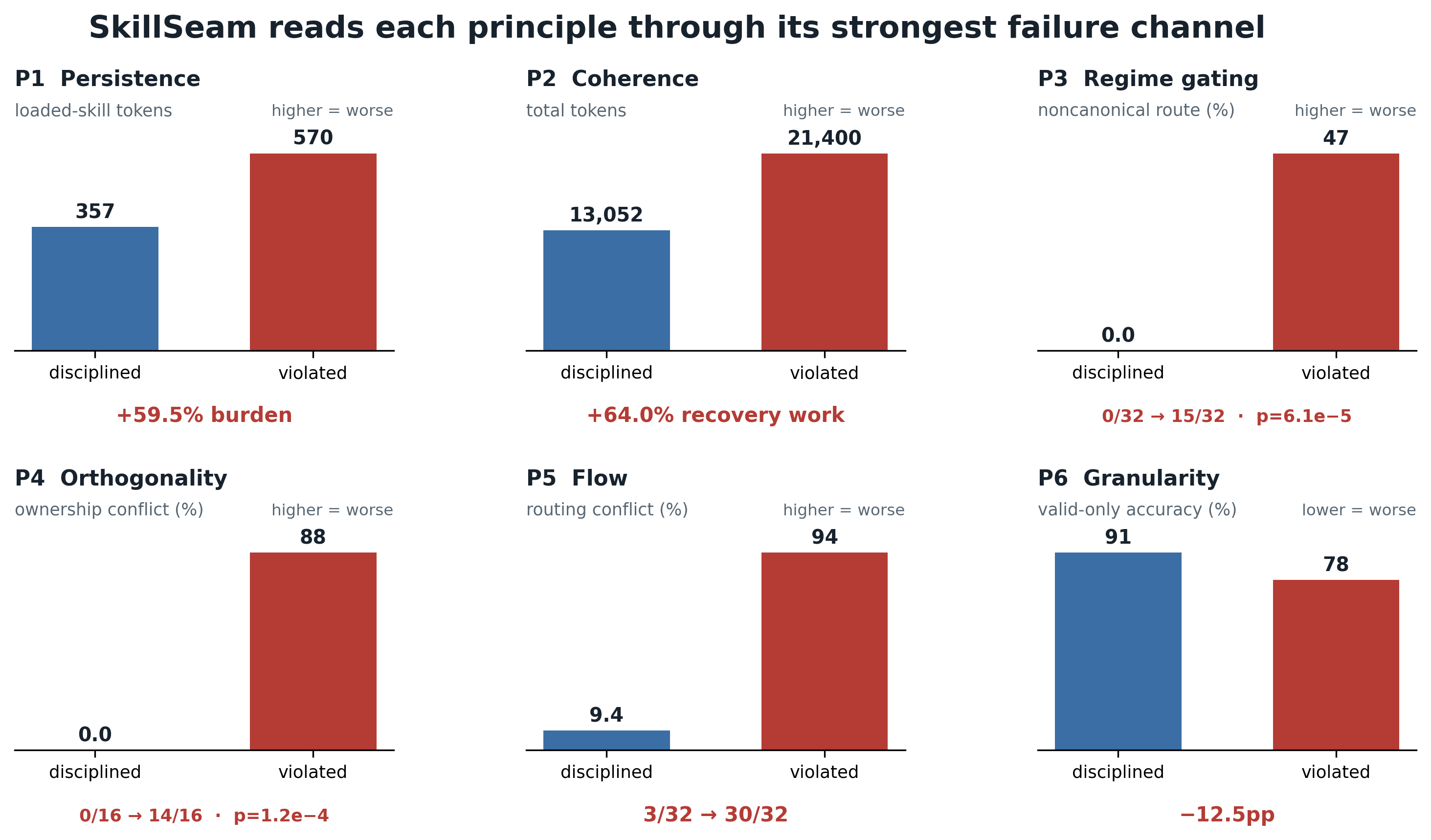}
\caption{\figcapprincipleevidence}
\label{fig:principle-evidence}
\end{figure}

\subsection{P1--P2: organization first appears as execution burden}
Flattening referenced detail into skill bodies (P1) does not hurt
accuracy: L1 scores 0.969 versus 0.903 at L0. Yet every activation now
carries more text. On the 31 tasks valid in both layers, mean
loaded-skill tokens rise from 357 to 570 ($+59.5\%$; median paired
increase 224 tokens). P1 is therefore an efficiency principle in these
data, not an accuracy principle.

Dangling one coherence anchor (P2) changes a different channel. Relative
to L1, mean total tokens rise from 13,052 to 21,400 ($+64.0\%$), while
accuracy moves from 0.969 to 0.938 ($-3.1$pp). The accuracy estimate is
small at this sample size; the more informative signature is extra
execution after the trace target becomes unreachable. Coherence protects
an agent from searching around a broken dependency.

\subsection{P3--P4: targeted routing probes expose partition failures}
On the original positive-task ladder, adding an alias (P3) and deleting
a lane boundary (P4) each produce a 0pp adjacent accuracy change. This
is informative rather than exculpatory: a positive task with one obvious
gold skill cannot expose competition between two plausible skills.

The P3 probe holds the collection at two skills and equalizes the
character length of the second skill. The baseline pairs the canonical
expense skill with an unrelated control; the treatment replaces only
that control with a same-regime synonym alias. Across 32 matched tasks,
noncanonical routing rises from $0/32$ to $15/32$ (exact McNemar
$p=6.10\times10^{-5}$), and same-intent canonical/synonym wording flips
the selected route in $0/16$ versus $8/16$ pairs. First-route entropy
rises from 0 to 0.989 bits, while double-reads rise from $0/32$ to
$12/32$. Regime gating therefore prevents routing fragmentation, not
merely extra files.

The P4 probe holds both arms at two skills with two sections per skill.
The baseline assigns report export and currency conversion to disjoint
owners; the treatment makes both skills claim both operations. On 16
balanced pre-execution ownership audits, reported duplicate ownership
rises from $0/16$ to $14/16$ (exact McNemar
$p=1.22\times10^{-4}$), and the wrong skill is read first in $8/16$
treatment sessions versus $0/16$ controls. This is a mechanism
diagnostic, not a natural-task accuracy estimate: on a separate
direct-execution slice, canonical-owner violations move only
$0/32\to3/32$ ($p=.25$). Orthogonality makes ownership decidable before
execution; ordinary success tasks often conceal that ambiguity.

\begin{table}[h]
\centering\scriptsize
\setlength{\tabcolsep}{3pt}
\caption{Illustrative released traces (English prompt glosses). Rows
show mechanism, not additional samples: aggregate estimates remain
those in Table~\ref{tab:principle-evidence}.}
\label{tab:route-traces}
\begin{tabular}{@{}p{0.7cm}p{1.5cm}p{4.0cm}p{4.0cm}p{4.3cm}@{}}
\toprule
 & Arm & Matched input & Observed read order & Routing decision \\
\midrule
P3 & disciplined & canonical: ``record one personal expense'' &
\texttt{expense-core} & canonical route \\
P3 & disciplined & synonym: ``log one personal purchase'' &
\texttt{expense-core} & same route \\
P3 & violated & canonical: ``record one personal expense'' &
\texttt{expense-core} & canonical route \\
P3 & violated & synonym: ``log one personal purchase'' &
\texttt{expense-alias $\rightarrow$ expense-core} & primary route flips to alias \\
\midrule
P4 & disciplined & export an expense report &
\texttt{report-lane $\rightarrow$ currency-lane} &
\texttt{OWNERSHIP\_CONFLICT: no} \\
P4 & violated & same request &
\texttt{currency-lane $\rightarrow$ report-lane} &
\texttt{OWNERSHIP\_CONFLICT: yes}; wrong owner read first \\
\bottomrule
\end{tabular}
\end{table}

Table~\ref{tab:route-traces} shows why the aggregate channels differ.
The P3 pair preserves semantics but changes the primary route only when
the alias exists. The P4 audit may inspect both files even in the
disciplined arm; the informative event is not the second read itself,
but whether those reads resolve to one owner. Under overlap, the same
request begins at the wrong lane and ends with an explicit conflict.

\subsection{P5: trigger flow produces the clearest routing failure}
Replacing observable yes/no triggers with bland descriptions causes a
routing collapse. Conflicts jump from $3/32$ at L4 to $30/32$ at L5;
27 paired tasks switch into conflict and none switch out (exact McNemar
$p=1.49\times10^{-8}$). Mean loaded-skill tokens rise from 552 to 2,061,
a 3.7$\times$ burden, while accuracy moves only $-3.1$pp. Accuracy alone
would nearly miss the strongest mechanism in the ladder.

A separate N16 arm provides a consistent warning: adding prose
\texttt{not-for} labels to the unmanaged collection moves valid-only
accuracy from 0.910 to 0.857 ($-5.3$pp). Those labels describe a boundary
but do not make the triggering condition more observable. The result is
reported as a corroborating estimate, not as a replacement for the
paired conflict evidence.

\subsection{P6: granularity reaches task correctness}
The final perturbation merges one too-thin skill into its parent and
splits one coarse skill behind blurred boundaries. Accuracy falls from
0.906 at L5 to 0.781 at L6 ($-12.5$pp), the largest adjacent accuracy
change in the ladder. Because L6 changes both sides of the granularity
rule, the estimate belongs to the combined mis-sizing intervention, not
to either merge or split alone.

\subsection{Supporting checks}
The larger mounting matrix answers secondary questions. First, the N4,
N16, and N64 sweeps show no monotone accuracy trend in either managed or
unmanaged layouts: within this range, file count does not explain the
organizational signatures above. Second, the N0 control gives the
expected sanity check that task-relevant skills matter at all (0.400
without skills versus 0.89--0.92 with a mounted collection). Full
supporting condition counts are reported in Appendix~\ref{sec:repro}.

\FloatBarrier

%% file: sections/principles.tex
\section{The SkillSeam Design Guide}
\label{sec:principles}

\begin{figure}[t]
\centering
\includegraphics[width=0.92\textwidth]{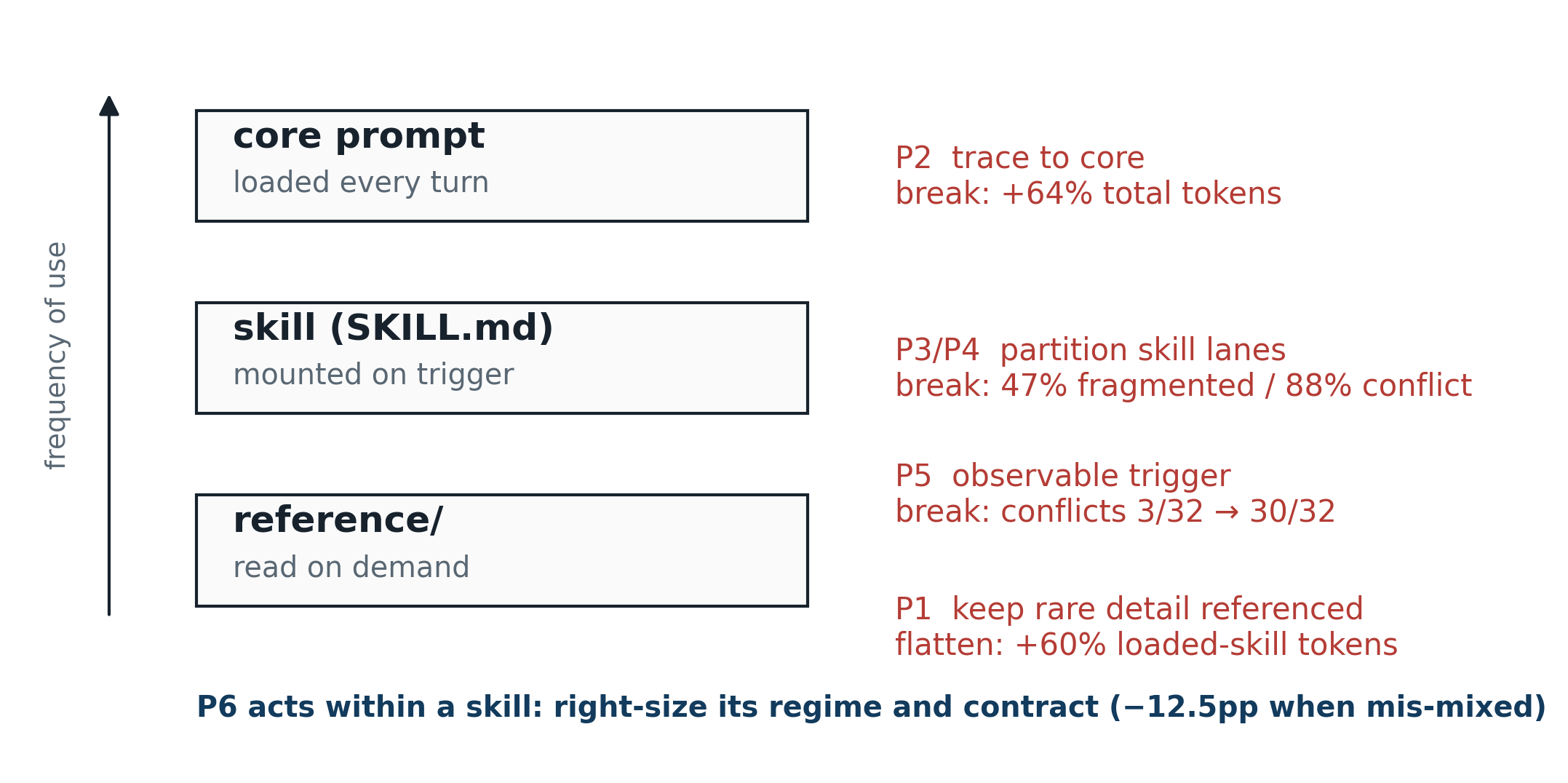}
\caption{\figcapconceptprinciples}
\label{fig:stack}
\end{figure}

The six SkillSeam principles treat skill authoring as organization
design. Skills are specialist roles; triggers are job descriptions;
routes are assignments; references are institutional memory; and tests
are accountable outcomes. The objective is not a larger roster, but an
organization whose failure channels remain observable.
Figure~\ref{fig:stack} locates the principles in the system;
Table~\ref{tab:checklist} is the operational checklist.

\begin{table*}[t]
\centering\small
\caption{The six-principle audit. Evidence status distinguishes direct
paired evidence from diagnostic-slice observations.}
\label{tab:checklist}
\begin{tabular}{@{}clp{4.5cm}p{3.4cm}p{3.8cm}@{}}
\toprule
 & Principle & Rule & Audit signal & Evidence in this paper \\
\midrule
P1 & Persistence gradient & Layer information by frequency of use & loaded-skill tokens & direct paired token burden \\
P2 & System coherence & Preserve each role's trace to system purpose & recovery effort after a broken trace & direct adjacent estimate \\
P3 & Regime gating & Create a role only for a new regime & route consistency under paraphrase & strong controlled paired evidence \\
P4 & Orthogonal coverage & Assign one accountable owner per lane & candidate ownership conflict & strong diagnostic paired evidence \\
P5 & Flow & Make entry conditions and handoffs observable & routing conflict and load burden & strong direct paired evidence \\
P6 & Granularity discipline & Right-size responsibilities & task-contract failure & direct adjacent estimate \\
\midrule
\multicolumn{5}{@{}l}{\emph{Substrate: self-contained verification~\cite{zhang2026skillspec}---every skill ships a runnable test.}} \\
\bottomrule
\end{tabular}
\end{table*}

\paragraph{P1 --- Build an information hierarchy.}
As an organization separates standing policy from team playbooks and
archived detail, put facts needed every turn in the core prompt, procedures behind skill
triggers, and rare detail in \texttt{reference/}. The test is not merely
whether the task succeeds; measure how much skill text each successful
session must carry. Flattening the hierarchy preserved accuracy here
but increased loaded-skill tokens by 59.5\%. A persistence audit should
therefore flag rarely used detail that rides with every activation.

\paragraph{P2 --- Give every role a traceable purpose.}
A role with no place in the organization chart becomes an orphan.
Likewise, every skill should implement a named system principle or capability.
Treat a missing trace target like a broken dependency: fail the build,
not the user session. When a trace is broken, inspect recovery effort
(tokens, turns, and tool calls) alongside correctness. The dangling
anchor in our ladder raised execution burden before it produced a large
accuracy signal.

\paragraph{P3 --- Create roles for regimes, not titles.}
A new role is justified by a distinct responsibility, not a new title;
a new skill is justified only by a new user--scenario regime. A second
name for the same behavior creates an assignment choice without adding
coverage. Test this principle with semantic hard negatives and count
noncanonical routes, paraphrase route flips, and same-semantic
multi-reads; generic success tasks can conceal the failure. If two
skills repeatedly load for one intent, merge them or
make their regimes observably different.

\paragraph{P4 --- Assign one accountable owner.}
For any routable system state, one skill should own the lane. As in a
well-defined responsibility matrix, advice may be shared but
accountability cannot be ambiguous. This is
not the same as writing exclusions in prose: the boundary must be
expressed in conditions the router can observe. Audit orthogonality on
candidate-ownership tasks and count cases where the router reports two
defensible owners or reads the wrong owner first.
P3 prevents duplicate regimes; P4 prevents neighboring regimes from
sharing the same boundary.

\paragraph{P5 --- Make assignments and handoffs observable.}
A usable job description tells colleagues when responsibility begins
and ends. Likewise, a trigger should answer a question about observable user intent before
it describes a method. Bland labels such as ``helps with scheduling''
force the model to infer the boundary at runtime. This principle has the
clearest direct signal in our data: conflict count and loaded-skill
burden both collapse when trigger faces are degraded. These routing
metrics should be first-class evaluation outputs, not debugging notes.

\paragraph{P6 --- Right-size each role.}
An overloaded role hides incompatible jobs; a fragmented role cannot
own an outcome. Split a skill when one file spans distinct regimes; fold a skill back
when it has no independent trigger, contract, or test. Count is not the
criterion---cohesion is. Granularity should be tested end-to-end because
mis-sizing changes which instructions and checks travel together. In
our ladder, the combined merge/split perturbation produces the largest
adjacent accuracy loss.

\paragraph{Verification substrate.}
Each rule becomes auditable only when a skill has an executable contract.
We adopt SkillSpec's view of skill correctness as specification
reasoning~\cite{zhang2026skillspec}: a self-check validates the expected
workspace effects independently of the model's final prose. Verification
is the floor beneath the six principles, not a seventh design claim.

%% file: sections/discussion.tex
\section{Discussion}
\label{sec:discussion}

\paragraph{A skill collection is a routing architecture.}
SkillSeam changes the unit of analysis from the individual instruction file
to the relationships among files. Once skills share an agent, quality
depends on where knowledge lives, which capability each file implements,
when each route becomes active, and what belongs together. These are
architecture questions. Treating them as prose quality misses the
system-level failures that appear only when otherwise reasonable skills
interact.

\paragraph{The metric must match the failure.}
The experiments explain why a single agent-accuracy score is inadequate.
A persistence violation first appears as excess context; an alias or
overlap first appears as route fragmentation or undecidable ownership;
a vague trigger appears as conflict and load amplification; a
granularity failure reaches the task contract. Positive-only benchmarks
can therefore certify a collection that fails under ambiguity. SkillSeam
offers a general evaluation pattern: derive the
failure condition from the design rule, construct the task slice that
activates it, and measure the closest observable before asking whether
aggregate accuracy moves.

\paragraph{Mechanism sensitivity and deployment incidence are different.}
A diagnostic task asks whether a failure can be activated under its
defining condition. A direct-execution task asks how often that
condition surfaces without assistance. P4 demonstrates the distinction:
ownership conflict is reported in $14/16$ audits, whereas only $3/32$
ordinary executions violate the canonical owner. The smaller number
does not refute the structural defect, and the larger number is not a
claim that 87.5\% of user tasks fail. Reporting both prevents two common
errors: dismissing a latent architecture problem because aggregate
accuracy is unchanged, and marketing a stress-test rate as production
prevalence.

\paragraph{Organization creates leading indicators.}
Several SkillSeam signals move before task correctness. P1 increases context
burden without reducing accuracy; P2 adds recovery work before a large
accuracy loss; P3 splits routes while preserving capability coverage;
and P4 makes ownership ambiguous even when either file contains a usable
procedure. These are useful release-gate metrics precisely because they
precede visible failure. They expose systems that still pass today's
tasks but have become harder to route, maintain, and extend.

\paragraph{Scale by audits, not accumulation.}
The N4--N64 sweep shows no monotone benefit from adding files in the
measured range. The practical bottleneck is not how many skills an agent
can mount, but whether the collection remains traceable, partitioned,
and routable. A team can apply SkillSeam as a release gate: measure loaded
skill burden, reject broken anchors, run alias and overlap suites, count
routing conflicts, and execute each skill's contract. This turns skill
maintenance from subjective cleanup into a repeatable systems test.

\paragraph{The six principles form a debugging order.}
The principles are not additive score components. They locate failures
at progressively later points in an execution path: P1 places knowledge,
P2 keeps it reachable, P3 decides whether a new route deserves to exist,
P4 assigns that route an owner, P5 makes its trigger observable, and P6
packages the resulting procedure at a testable grain. This ordering
gives practitioners a triage strategy. Fixing a downstream prompt cannot
repair an unreachable trace, and tuning task logic cannot make duplicate
owners disjoint. Start at the earliest violated relationship, then rerun
the mechanism-specific probe.

%% file: sections/related.tex
\section{Related Work}
\label{sec:related}

\paragraph{Procedural knowledge and skill libraries.}
Agents increasingly externalize reusable behavior. ReAct couples
reasoning and action~\cite{yao2023react}; Voyager stores executable
skills~\cite{wang2023voyager}; Agent Workflow Memory and ExpeL distill
trajectories into reusable procedures~\cite{wang2024awm,zhao2024expel};
and programmatic skill induction and skill networks add verification
and library refactoring~\cite{wang2025asi,shi2026psn}. Memory systems
similarly organize persistent experience~\cite{park2023generative,packer2023memgpt,zhang2024memsurvey}.
This line asks what agents should retain and retrieve. SkillSeam instead
holds the procedural layer fixed enough to study how relationships
among skills create routing and execution failures.

\paragraph{Context and tool routing.}
RAG and context-engineering research studies which information enters a
model's context and how position, length, and retrieval affect
performance~\cite{lewis2020rag,liu2023lostmiddle,gao2023ragsurvey,mei2025contextengsurvey}.
Tool and skill guidance recognizes the same operational pressure:
context is finite, tools need functional boundaries, and detailed
instructions should load on demand~\cite{anthropic2025contexteng,anthropic2025writingtools,anthropic2025claudecode}.
Those works motivate individual authoring choices. SkillSeam turns six such
collection-level choices into perturbations with distinct observables:
context burden, broken traces, multi-reads, routing conflicts, and task
failure.

\paragraph{Agent evaluation and specification.}
Reproducible evaluation frameworks emphasize fixed conditions and
inspectable task contracts~\cite{liang2022helm,gao2023lmeval}; agent
benchmarks extend that discipline to repositories, browsers, tools, and
computer use~\cite{jimenez2024swebench,zhou2023webarena,yao2024taubench,xie2024osworld}.
SkillSeam applies controlled evaluation to the skill collection itself: one
organizational property becomes the treatment, and the task slice is
chosen to expose its predicted failure. The six principle semantics
come from the System-of-Skills guide~\cite{kr33skills}; the executable
verification substrate adopts SkillSpec's view of skill correctness as
specification reasoning~\cite{zhang2026skillspec}. The contribution is
not another skill format, but a way to measure whether a collection's
design rules survive contact with agent behavior.

%% file: sections/limits.tex
\section{Limitations}
\label{sec:limits}

The principle ladder uses one model, one seed, and one suite family;
other systems may change the magnitudes. P4's strongest result is an
ownership-audit diagnostic, not a natural-task failure rate; the
separate direct-execution slice is smaller. Token fields are parser
estimates, and cumulative ladder effects are adjacent signatures rather
than independent constants. Denominators and exclusions appear in
Appendix~\ref{sec:repro}.

%% file: sections/conclusion.tex
\section{Conclusion}
\label{sec:conclusion}

A roster becomes an organization when roles, information, ownership,
and handoffs become explicit. A skill collection becomes a system by
the same logic. SkillSeam makes those relationships measurable: each of six design
principles is paired with a failure mechanism, a targeted perturbation,
and the observable best suited to reveal it. Four perturbations expose
clear context, recovery, routing, and accuracy signatures. Controlled
follow-ups complete the map: a synonymous alias creates noncanonical
routes in $15/32$ tasks, while overlapping lanes produce reported
ownership conflict in $14/16$ audits. Together with a $27/32$ conflict
increase for trigger flow and a $-12.5$pp granularity estimate, the
results show why no single accuracy score can price skill organization.
The lesson is not to add fewer or more skills, but to organize the
capability already present: layer its information, preserve its traces,
gate new roles, assign one owner, expose handoffs, and right-size each
responsibility. Better agents need skill collections whose seams can be
inspected; SkillSeam supplies that audit.

%% file: sections/appendix-repro.tex
\section{Reproducibility and Supporting Results}
\label{sec:repro}

\subsection{Supporting mounting matrix}
Table~\ref{tab:supporting-matrix} reports the conditions removed from
the main reading path. They establish that skills are consequential and
that file count alone does not explain the SkillSeam signatures; they are not
estimates of the six principle effects.

\begin{table}[h]
\centering\footnotesize
\caption{Supporting conditions. Accuracy is valid-only; valid/session
counts expose every exclusion.}
\label{tab:supporting-matrix}
\begin{tabular}{@{}llrrl@{}}
\toprule
Model/leg & Condition & Sessions & Valid & Accuracy \\
\midrule
glm matrix & managed N4 / N16 / N64 & 300 & 297 & .909 / .918 / .910 \\
glm matrix & unmanaged N4 / N16 / N64 & 300 & 299 & .890 / .910 / .909 \\
glm control & none N0 & 100 & 100 & .400 \\
glm ablation & managed minus refsink & 100 & 99 & .929 \\
glm ablation & unmanaged plus \texttt{not-for} & 100 & 98 & .857 \\
deepseek breadth & managed / unmanaged N16 & 200 & 168 & .855 / .812 \\
\bottomrule
\end{tabular}
\end{table}

\subsection{Analysis notes}
The main ladder contains 32 sessions per layer and one invalid L0
session. Continuous metrics use tasks valid in both adjacent layers.
P1's paired loaded-skill token comparison therefore uses 31 tasks
(357.4 to 570.0 mean tokens; median paired increase 224). P5's conflict
comparison uses all 32 matched tasks: 27 switch into conflict and none
switch out, giving a two-sided exact McNemar
$p=1.49\times10^{-8}$. Other adjacent accuracy contrasts are reported
as effect estimates because their exact paired tests do not reject at
the pre-registered corrected threshold.

An initial 30-session-per-arm P3/P4 slice failed to activate either
designated pair; we retain that null result in the released analysis
rather than treating unrelated multi-reads as evidence. The redesigned
GLM-5.3-Flash probes isolate the mechanisms. P3 has 32 valid matched
tasks in two equal-sized, equal-character-budget collections; 15 tasks
switch only into noncanonical routing ($p=6.10\times10^{-5}$). P4 has
16 valid balanced candidate-audit pairs with equal skill and section
counts; 14 switch only into reported duplicate ownership
($p=1.22\times10^{-4}$). A separate 32-pair direct-execution P4 slice
moves $0\to3$ ($p=.25$), so the paper labels the stronger P4 result as
a diagnostic rather than an ecological failure rate.

\subsection{Probe robustness by intent}
Table~\ref{tab:probe-strata} checks whether the headline effects are
carried by one wording family. P3's treatment produces noncanonical
routes in every tested operation: recording, aggregation, budget
checking, and budget setting. The magnitude varies from $2/8$ to $6/8$,
which is expected because descriptions and prompts differ in lexical
proximity, but no intent has a zero treatment count. All corresponding
baseline strata remain at zero.

\begin{table}[h]
\centering\small
\caption{Controlled-probe outcomes by intent. Counts use only sessions
valid in both arms.}
\label{tab:probe-strata}
\begin{tabular}{@{}llrrl@{}}
\toprule
Probe & Intent & Disciplined & Violated & Measured event \\
\midrule
P3 & record expense & $0/8$ & $4/8$ & noncanonical route \\
P3 & aggregate expenses & $0/8$ & $2/8$ & noncanonical route \\
P3 & check budget & $0/8$ & $3/8$ & noncanonical route \\
P3 & set budget & $0/8$ & $6/8$ & noncanonical route \\
\midrule
P4 & currency conversion & $0/8$ & $7/8$ & ownership conflict \\
P4 & report export & $0/8$ & $7/8$ & ownership conflict \\
\bottomrule
\end{tabular}
\end{table}

P4 is exactly balanced across its two target directions: both currency
conversion and report export produce $7/8$ reported conflicts under
overlap and none under disjoint ownership. This symmetry matters because
an order artifact would more naturally favor the skill listed or read
first. Instead, treatment sessions read the wrong owner first in
$8/16$ cases while the explicit conflict decision rises in both
directions. The result therefore reflects duplicated ownership rather
than one anomalous skill name.

\subsection{Probe execution and integrity}
Each probe task runs in a fresh agent session with only the two
condition-specific skills mounted. The generator enforces matched task
IDs, a byte-identical canonical P3 skill, equal P3 second-file character
counts, equal P4 skill counts, and equal P4 section counts. Offline tests
check those invariants before any model request. Analysis intersects
valid task IDs across arms, deduplicates skill reads by first occurrence,
and computes exact two-sided McNemar tests from paired binary outcomes.

The released records include system prompts, ordered skill reads, final
decisions, token usage, and condition manifests. The OpenRouter model ID
is \texttt{z-ai/glm-5.3-flash}; no API credential is stored in an
artifact. Across smoke, completed, and retained diagnostic runs, the
harness recorded \$0.0742 in model cost. Cost is reported for
reproducibility only and is not an outcome of the six-principle study.

\subsection{Falsification checks}
SkillSeam treats each measured signature as support only after checking the
most immediate competing explanation. Table~\ref{tab:falsification}
summarizes those checks. The purpose is not to claim that every possible
confound has been removed; it is to keep the conclusion no broader than
the contrast permits.

\begin{table}[h]
\centering\footnotesize
\caption{Claim-level falsification checks. Each row identifies the
nearest alternative explanation and the design feature used to test it.}
\label{tab:falsification}
\begin{tabular}{@{}cp{5.7cm}p{8.4cm}@{}}
\toprule
 & Nearest alternative & Check built into SkillSeam \\
\midrule
P1 & Harder tasks caused more context use & Same-task valid pairs; only referenced detail is moved into the loaded body \\
P2 & Token growth is ordinary sampling noise & Same tasks and model settings; the sole new defect is one unreachable named anchor \\
P3 & More files or more characters created the split & Two skills in both arms; exact second-file character match; canonical skill byte-identical \\
P4 & One skill name or listing order caused the conflict & Same named lanes in both arms; both report and currency directions yield $7/8$ conflicts \\
P5 & Accuracy alone captures the trigger defect & Target routing conflicts move $27$ pairs in one direction while accuracy changes only $-3.1$pp \\
P6 & The result identifies merge and split separately & Claim is explicitly restricted to the combined mis-sizing intervention \\
\bottomrule
\end{tabular}
\end{table}

The checks also determine the appropriate language of each conclusion.
P1 and P2 support burden claims, not universal accuracy losses. P3
supports route fragmentation after a same-regime alias is introduced,
not the proposition that every alias causes failure on every prompt.
P4 supports undecidable ownership under an explicit audit, while its
direct-execution rate remains separately visible. P5 supports a routing
collapse because the designated event, not a generic token count,
changes in 27 matched pairs. P6 supports the practical granularity rule
as a combined intervention; separating over- and under-granulation would
require another factorial experiment. This claim discipline lets the
paper emphasize strong effects without turning a sensitive diagnostic
into a broader statement than the data warrant.

\subsection{Code, variants, and rollups}
Every main-text number is regenerated by
\path{06-paper/analysis/compute_paper_stats.py}; its tests are in
\path{06-paper/analysis/test_compute_paper_stats.py}. Source data
are the JSONL rollups in \path{05-analysis/}, including
\path{rollup-ladder-p3.json}, \path{rollup-ladder-p4.json}, the
three paid-matrix rollups, the breadth rollup, and
\path{p3-regime-probe-glm53f.json},
\path{p4-orthogonality-audit-glm53f.json}, and
\path{p4-direct-probe-glm53f.json}.

The frozen scheduler, parser, and offline gate are
\path{02-harness/run_matrix.py}, \path{02-harness/parser.py}, and
\path{02-harness/tests/run_all.py}. Condition manifests are
\path{02-harness/conditions.json},
\path{conditions-ladder.json}, and
\path{conditions-ladder-p4.json}. The controlled follow-up generator,
task suites, collections, conditions, and analyzer are under
\path{03-suites/principle-probes/}. Byte-hashed L0--L6 variants and
their generator live under
\path{03-suites/l-skill-variants/}; each snapshot includes a layer
manifest and previous-layer diff. The fixed generator seed is 24301.

\subsection{Companion agent skill}
The SkillSeam principles are also distributed as an agent-readable skill in
the public System-of-Skills repository~\cite{kr33skills}:
\url{https://github.com/X32Studio/best-practice-for-skills-system}.
The repository includes English and Chinese guides, principle pointers,
and an installable \texttt{skill/} directory. The paper supplies the
measurement evidence; the companion skill supplies the operational
interface an agent can apply while creating or reviewing a collection.

%% file: references.bib
@misc{anthropic2025skills,
  title        = {Agent Skills: Best Practices},
  author       = {{Anthropic}},
  year         = {2025},
  howpublished = {\url{https://docs.anthropic.com/en/docs/agents-and-tools/agent-skills/best-practices}},
  note         = {Accessed 2026-09 for principle-level comparison}
}

@misc{kr33skills,
  title        = {System of Skills: How to Organize a Set of Agent Skills Into a Coherent System},
  author       = {k-r33 (X32 Studio)},
  year         = {2026},
  howpublished = {\url{https://github.com/X32Studio/best-practice-for-skills-system}},
  note         = {Upstream practitioner guide and installable agent skill. This paper evaluates six collection-organization principles and treats self-contained verification as their substrate}
}

@misc{zhang2026skillspec,
  title         = {{SkillSpec}: Intent-Masked Specification Reasoning for Agent Skill Correctness},
  author        = {Zhang, Yizhuo and Kang, Bo and Yang, Yi and Duan, Zhiyu and Ye, Zhouteng and Yang, Shunkun},
  year          = {2026},
  eprint        = {2609.06052},
  archivePrefix = {arXiv},
  primaryClass  = {cs.SE},
  note          = {Source of the self-contained-verification semantics (skill correctness as specification reasoning); cited with credit}
}

@article{packer2023memgpt,
  title   = {{MemGPT}: Towards {LLMs} as Operating Systems},
  author  = {Packer, Charles and Wooders, Sarah and Lin, Kevin and Fang, Vivian and Patil, Shishir G. and Zhang, Tianjun and Gonzalez, Joseph E.},
  journal = {arXiv preprint arXiv:2310.08560},
  year    = {2023}
}

@article{wang2024awm,
  title   = {Agent Workflow Memory},
  author  = {Wang, Zora Zhiruo and Mao, Jiayuan and Fried, Daniel and Neubig, Graham},
  journal = {arXiv preprint arXiv:2409.07429},
  year    = {2024}
}

@article{liang2022helm,
  title   = {Holistic Evaluation of Language Models},
  author  = {Liang, Percy and Bommasani, Rishi and Lee, Tony and others},
  journal = {arXiv preprint arXiv:2211.09110},
  year    = {2022},
  note    = {Methodological template for condition-matrix, pre-registered LLM evaluation}
}

@article{wang2023voyager,
  title   = {Voyager: An Open-Ended Embodied Agent with Large Language Models},
  author  = {Wang, Guanzhi and Xie, Yuqi and Jiang, Yunfan and Mandlekar, Ajay and Xiao, Chaowei and Zhu, Yuke and Fan, Linxi and Anandkumar, Anima},
  journal = {arXiv preprint arXiv:2305.16291},
  year    = {2023}
}

@article{park2023generative,
  title   = {Generative Agents: Interactive Simulacra of Human Behavior},
  author  = {Park, Joon Sung and O'Brien, Joseph C. and Cai, Carrie J. and Morris, Meredith Ringel and Liang, Percy and Bernstein, Michael S.},
  journal = {arXiv preprint arXiv:2304.03442},
  year    = {2023}
}

@article{yao2023react,
  title   = {{ReAct}: Synergizing Reasoning and Acting in Language Models},
  author  = {Yao, Shunyu and Zhao, Jeffrey and Yu, Dian and Du, Nan and Shafran, Izhak and Narasimhan, Karthik and Cao, Yuan},
  journal = {arXiv preprint arXiv:2210.03629},
  year    = {2023},
  note    = {ICLR 2023}
}

@article{zhang2024memsurvey,
  title   = {A Survey on the Memory Mechanism of Large Language Model based Agents},
  author  = {Zhang, Zeyu and Bo, Xiaohe and Ma, Chen and Li, Rui and Chen, Xu and Dai, Quanyu and Zhu, Jieming and Dong, Zhenhua and Wen, Ji-Rong},
  journal = {arXiv preprint arXiv:2404.13501},
  year    = {2024},
  note    = {(arXiv:2404.13501, not the look-alike 2304.13501 which is an unrelated networking paper)}
}

@article{zhao2024expel,
  title   = {{ExpeL}: {LLM} Agents Are Experiential Learners},
  author  = {Zhao, Andrew and Huang, Daniel and Xu, Quentin and Lin, Matthieu and Liu, Yong-Jin and Huang, Gao},
  journal = {arXiv preprint arXiv:2308.10144},
  year    = {2024},
  note    = {AAAI 2024}
}

@article{wang2025asi,
  title   = {Inducing Programmatic Skills for Agentic Tasks},
  author  = {Wang, Zora Zhiruo and Gandhi, Apurva and Neubig, Graham and Fried, Daniel},
  journal = {arXiv preprint arXiv:2504.06821},
  year    = {2025}
}

@article{shi2026psn,
  title   = {Evolving Programmatic Skill Networks},
  author  = {Shi, Haochen and Yuan, Xingdi and Liu, Bang},
  journal = {arXiv preprint arXiv:2601.03509},
  year    = {2026},
  note    = {Skill libraries whose structure is refactored under rollback validation; organization of the skill library as a first-class object}
}

@article{lewis2020rag,
  title   = {Retrieval-Augmented Generation for Knowledge-Intensive {NLP} Tasks},
  author  = {Lewis, Patrick and Perez, Ethan and Piktus, Aleksandra and Petroni, Fabio and Karpukhin, Vladimir and Goyal, Naman and K{\"u}ttler, Heinrich and Lewis, Mike and Yih, Wen-tau and Rockt{\"a}schel, Tim and Riedel, Sebastian and Kiela, Douwe},
  journal = {arXiv preprint arXiv:2005.11401},
  year    = {2020},
  note    = {NeurIPS 2020}
}

@article{liu2023lostmiddle,
  title   = {Lost in the Middle: How Language Models Use Long Contexts},
  author  = {Liu, Nelson F. and Lin, Kevin and Hewitt, John and Paranjape, Ashwin and Bevilacqua, Michele and Petroni, Fabio and Liang, Percy},
  journal = {arXiv preprint arXiv:2307.03172},
  year    = {2023},
  note    = {TACL 2023}
}

@article{gao2023ragsurvey,
  title   = {Retrieval-Augmented Generation for Large Language Models: A Survey},
  author  = {Gao, Yunfan and Xiong, Yun and Gao, Xinyu and Jia, Kangxiang and Pan, Jinliu and Bi, Yuxi and Dai, Yi and Sun, Jiawei and Wang, Meng and Wang, Haofen},
  journal = {arXiv preprint arXiv:2312.10997},
  year    = {2023}
}

@article{mei2025contextengsurvey,
  title   = {A Survey of Context Engineering for Large Language Models},
  author  = {Mei, Lingrui and Yao, Jiayu and Ge, Yuyao and Wang, Yiwei and Bi, Baolong and Cai, Yujun and Liu, Jiazhi and Li, Mingyu and Li, Zhong-Zhi and Zhang, Duzhen and Zhou, Chenlin and Mao, Jiayi and Xia, Tianze and Guo, Jiafeng and Liu, Shenghua},
  journal = {arXiv preprint arXiv:2507.13334},
  year    = {2025}
}

@misc{anthropic2025contexteng,
  title        = {Effective Context Engineering for {AI} Agents},
  author       = {{Anthropic Applied AI team (Rajasekaran, P.; Dixon, E.; Ryan, C.; Hadfield, J.)}},
  year         = {2025},
  howpublished = {\url{https://www.anthropic.com/engineering/effective-context-engineering-for-ai-agents}},
  note         = {Published 2025-09-29; page fetched and title matched 2026-09-10}
}

@misc{anthropic2025claudecode,
  title        = {Best Practices for {Claude} Code},
  author       = {{Anthropic}},
  year         = {2025},
  howpublished = {\url{https://code.claude.com/docs/en/best-practices}},
  note         = {Canonical URL (redirects from /engineering/claude-code-best-practices); page fetched, title matched 2026-09-10. Skills-as-CLAUDE.md-vs-SKILL.md placement guidance}
}

@article{jimenez2024swebench,
  title   = {{SWE-bench}: Can Language Models Resolve Real-World {GitHub} Issues?},
  author  = {Jimenez, Carlos E. and Yang, John and Wettig, Alexander and Yao, Shunyu and Pei, Kexin and Press, Ofir and Narasimhan, Karthik},
  journal = {arXiv preprint arXiv:2310.06770},
  year    = {2024},
  note    = {ICLR 2024}
}

@article{zhou2023webarena,
  title   = {{WebArena}: A Realistic Web Environment for Building Autonomous Agents},
  author  = {Zhou, Shuyan and Xu, Frank F. and Zhu, Hao and Zhou, Xuhui and Lo, Robert and Sridhar, Abishek and Cheng, Xianyi and Ou, Tianyue and Bisk, Yonatan and Fried, Daniel and Alon, Uri and Neubig, Graham},
  journal = {arXiv preprint arXiv:2307.13854},
  year    = {2023}
}

@article{xie2024osworld,
  title   = {{OSWorld}: Benchmarking Multimodal Agents for Open-Ended Tasks in Real Computer Environments},
  author  = {Xie, Tianbao and Zhang, Danyang and Chen, Jixuan and Li, Xiaochuan and Zhao, Siheng and Cao, Ruisheng and Hua, Toh Jing and Cheng, Zhoujun and Shin, Dongchan and Lei, Fangyu and others},
  journal = {arXiv preprint arXiv:2404.07972},
  year    = {2024}
}

@article{yao2024taubench,
  title   = {$\tau$-bench: A Benchmark for Tool-Agent-User Interaction in Real-World Domains},
  author  = {Yao, Shunyu and Shinn, Noah and Razavi, Pedram and Narasimhan, Karthik},
  journal = {arXiv preprint arXiv:2406.12045},
  year    = {2024}
}

@misc{gao2023lmeval,
  title        = {EleutherAI/lm-evaluation-harness},
  author       = {{EleutherAI (Gao, Leo; Tow, Jonathan; et al.)}},
  year         = {2023},
  howpublished = {Zenodo, DOI \url{10.5281/zenodo.10256836}, \url{https://zenodo.org/records/10256836}},
  note         = {Framework for few-shot LM evaluation; record fetched and title matched 2026-09-10}
}

@misc{anthropic2025writingtools,
  title        = {Writing Effective Tools for {AI} Agents --- With {AI} Agents},
  author       = {{Anthropic (Aizawa, Ken) and colleagues}},
  year         = {2025},
  howpublished = {\url{https://www.anthropic.com/engineering/writing-tools-for-agents}},
  note         = {Published 2025-09-11; page fetched, title matched 2026-09-10; "more tools don't always lead to better outcomes" and namespacing/boundary guidance}
}
